\documentclass[american]{article}
\usepackage[T1]{fontenc}
\usepackage[utf8]{inputenc}
\usepackage[american]{babel}
\usepackage{geometry}
\usepackage{amsmath}
\usepackage{amssymb}
\usepackage{graphicx}
\usepackage{wasysym}
\providecommand{\tabularnewline}{\\}

\begin{document}
\title{Implications of the muon anomalous magnetic moment in a Doublet Left-Right
Symmetric Model}
\author{M. Zeleny-Mora$^{1}$\thanks{moiseszeleny@gmail.com}, R. Gaitán-Lozano$^{1}$\thanks{regaitan@gmail.com},
R. Martinez$^{2}$\thanks{remartinezm@unal.edu.co}\\
 {\small$^{1}$Departamento de Física, FES-Cuautitlán, UNAM, C.P.
54770, Estado de México, México.}\\
{\small$^{2}$Departamento de Física, Universidad Nacional de Colombia,
K. 45 No. 26-85, Bogotá, Colombia.}}
\maketitle
\begin{abstract}
We compute the complete set of one-loop contributions to the muon
anomalous magnetic moment, $a_{\mu}=(g-2)_{\mu}/2$, in the Doublet
Left-Right Symmetric Model (DLRSM), based on the gauge group
$SU(2)_{L}\otimes SU(2)_{R}\otimes U(1)_{B-L}$ with neutrino masses
generated via the inverse seesaw (ISS) mechanism. We evaluate all four
one-loop topologies---VFF, SFF, FVV, and FSS---arising from the
extended gauge bosons ($W^{\prime}$, $Z^{\prime}$), the new scalar
sector ($H_{3}^{0}$, $A_{1}^{0}$, $H_{R}^{\pm}$, $H_{L}^{\pm}$),
and the heavy neutrino spectrum generated by the ISS mechanism,
using the Casas--Ibarra parametrization to express the neutrino
mixing in terms of physical observables. Imposing
the experimental bound on $\Delta a_{\mu}$, we establish that
$v_{R}\lesssim1$ TeV is excluded, implying lower bounds
$m_{W^{\prime}}\gtrsim325$ GeV, $m_{Z^{\prime}}\gtrsim385$ GeV,
and $m_{N}\gtrsim700$ GeV under the manifest left-right symmetry
condition $g_{R}=g_{L}$. Relaxing this condition to $g_{R}\neq g_{L}$
strengthens the gauge boson bounds to $m_{W^{\prime}}\gtrsim1625$ GeV
and $m_{Z^{\prime}}\gtrsim1650$ GeV.
\end{abstract}

\section{Introduction}

The Standard Model (SM) of particle physics has been remarkably successful
in describing fundamental interactions at the electroweak scale. Nonetheless,
several open questions remain, including the origin of neutrino masses,
the nature of parity violation, and the existence of lepton flavor
violation (LFV). In fact, neutrino oscillation experiments confirm
that neutrinos have nonzero masses, and therefore a new mass generation
mechanism is needed to describe this small scale. An accepted explanation
is the type-I seesaw mechanism which introduces right-handed neutrinos
$\nu_{R}$ and heavy Majorana neutrinos with mass matrix $M_{R}$,
for three right-handed neutrino generations one obtains a $6\times6$
full neutrino mass matrix $\mathcal{M}_{\nu}$. In the limit $|m_{D}|\ll|M_{R}|$,
the light neutrino mass matrix $\mathcal{M}_{\text{light}}$ is approximately
\[
\mathcal{M}_{\text{light}}\approx-m_{D}^{\top}M_{R}^{-1}m_{D}.
\]

where $m_{D}$ is the Dirac mass matrix. Typically, this requires
$M_{R}\approx10^{14}$ GeV, a scale beyond current experimental reach.
The inverse seesaw (ISS) mechanism offers an alternative by introducing
three pairs of fermionic singlets $\left(N_{R},S\right)$. In addition
to $M_{R}$ a new small Majorana mass matrix $\mu_{S}$ for the singlets
$S$ is included. Assuming the hierarchy $|\mu_{S}|\ll|m_{D}|\ll|M_{R}|$,
the light neutrino mass matrix is approximated as follows
\[
\mathcal{M}_{\text{light}}\approx m_{D}^{\top}M_{R}^{-1}\mu_{S}\left(M_{R}^{\top}\right)^{-1}m_{D},
\]
allowing right-handed neutrinos to reside at the TeV scale, potentially
within reach of current collider experiments.

The Left-Right Symmetric Model (LRSM) offers a compelling framework
to address parity violation by extending the SM gauge group to 
\[
SU\left(3\right)_{C}\otimes SU\left(2\right)_{L}\otimes SU\left(2\right)_{R}\otimes U\left(1\right)_{B-L},
\]

restoring left-right symmetry at higher energies \cite{Pati:1974yy,Mohapatra:1974gc,Mohapatra:1974hk,Senjanovic:1975rk}.
In contrast to the canonical version based on triplet scalar fields
and neutrino masses described by the seesaw type I or II, the Doublet
Left-Right Symmetric Model (DLRSM) introduces a scalar sector consisting
of a bidoublet $\Phi$ and two doublets $\chi_{L}$ and $\chi_{R}$,
which simplifies the scalar potential\cite{SENJANOVIC1979334} and
allows the incorporation of ISS mechanism to generate light neutrino
masses with right-handed neutrinos at TeV scale \cite{Mohapatra:1986bd,Dev:2012sg}.
At hadron colliders, the canonical production signature of such TeV-scale
states is the Keung-{}-Senjanovic (KS) process, $pp\to W^{\prime\pm}\to\ell^{\pm}N\to\ell^{\pm}\ell^{\pm}jj$,
in which an on-shell $W^{\prime}$ decays to a charged lepton and
a heavy Majorana neutrino, yielding a same-sign dilepton plus dijet
final state, this constitutes a distinctive signature of lepton-number
violation at colliders~\cite{Keung:1983uu}.

The symmetry-breaking scale $v_{R}$ of the LRSM is already tightly
constrained from the quark flavor sector. Indirect CP violation in
the neutral kaon system via the chromomagnetic dipole operator, which
receives a GIM-enhanced coefficient of order $10^{5}\zeta$ relative
to the SM contribution, implies a lower bound $M_{W^{\prime}}\gtrsim2.8-3$
TeV on the right-handed gauge boson mass for phenomenologically viable
phase configurations~\cite{Bertolini:2012gu}. A comprehensive analysis
of $\Delta F=2$ transitions in the $K$, $B_{d}$ and $B_{s}$ meson
systems including one-loop self-energy and vertex renormalization
diagrams required for a gauge-independent result shows that $B$-meson
oscillation data place constraints competitive with those from kaon
physics, yielding an absolute lower bound $M_{W^{\prime}}\gtrsim2.9-3.1$
TeV at $95\%$ CL~\cite{Bertolini:2014sua}. Complementary to these
quark-sector bounds, precision lepton observables such as the anomalous
magnetic moment of the muon, $a_{\mu}=(g-2)_{\mu}/2$, probe the LRSM
parameter space through a different set of loop topologies.

The present world average synthesizes two decades of precision measurements.
The BNL E821 experiment provided the first high-precision determination,
establishing a long-standing discrepancy of approximately $3.7\sigma$
with the then-current SM prediction~\cite{Bennett:2006fi}. The Fermilab
E989 experiment subsequently confirmed and sharpened this result in
three successive campaigns: Run $1$ at $0.46$ ppm~\cite{Abi:2021gix},
Runs $2-3$ at $0.20$ ppm~\cite{Aguillard:2023moo}, and the final
combination of all six runs at an unprecedented $127$ ppb~\cite{Muong-2:2025xyk}
a fourfold improvement over BNL and the most precise measurement of
$a_{\mu}$ ever achieved.

Recently, experimental searches for the anomalous magnetic moment
of the muon $a_{\mu}=\left(g-2\right)_{\mu}/2$ performed by Muon
$g-2$ Experiment at Fermi National Accelerator Laboratory (FNAL)
has contributed to improve the world-average determination of \cite{Muong-2:2025xyk}
\[
a_{\mu}^{\text{exp}}=1165920715\left(145\right)\times10^{-12}.
\]
with an unprecedented precision of 127 ppb. The SM prediction of $a_{\mu}$
has recently been refined, with lattice-QCD providing a consolidated
determination of the leading-order hadronic vacuum polarization at
$0.9\%$ precision. Incorporating this result yields \cite{ALIBERTI20251},
\[
a_{\mu}^{\text{SM}}=116592033\left(62\right)\times10^{-11},
\]
which agrees with the current experimental world average within uncertainties.
Together, these results establish 
\[
\Delta a_{\mu}=a_{\mu}^{\text{exp}}-a_{\mu}^{\text{SM}}=\left(38\pm63\right)\times10^{-11},
\]
which implies no tension with the SM prediction at the current level
of precision \cite{ALIBERTI20251}. For Beyond the Standard Model
(BSM) scenarios the new contributions to $a_{\mu}$ should be positive
or negative but with a sufficiently small magnitude \cite{Athron:2025ets}. 

Motivated by these constraints, the implications of $\Delta a_{\mu}$
have been investigated in a broad class of gauge extensions of the
SM. In the context of $331$ models, which extend the electroweak
gauge group to $SU(3)_{C}\otimes SU(3)_{L}\otimes U(1)$, a systematic
examination of all one\nobreakdash-loop topologies with neutral and
charged scalar fields, singly and doubly charged gauge bosons, and
neutral gauge bosons has been performed, showing that complementary
electroweak, dark-matter, and collider constraints strongly restrict
the symmetry breaking scale $v_{\chi}$ of these models~\cite{PhysRevD.90.113011}.
In extensions with non-universal Abelian symmetries and an inverse
seesaw neutrino sector, it has been found that the dominant positive
contribution to $\Delta a_{\mu}$ originates from the SM $W^{+}$
boson interacting with exotic heavy Majorana neutrinos at one loop,
while contributions from the extended neutral gauge sector and from
the scalar sector are subleading or negative~\cite{PhysRevD.108.095040}.

In this work, we compute all one-loop contributions to $a_{\mu}$
within the DLRSM, covering the four topologies VFF, SFF, FVV, and
FSS arising from the extended neutral and charged gauge bosons ($Z^{\prime}$
and $W^{\prime}$), the new scalar sector ($H_{3}^{0}$, $A_{1}^{0}$,
$H_{R}^{\pm}$, $H_{L}^{\pm}$), and the heavy neutrino spectrum generated
by the ISS mechanism. Using the experimental constraint on $\Delta a_{\mu}$,
we derive bounds on the right-handed symmetry-breaking scale $v_{R}$
and the associated heavy-state mass spectrum. The paper is organized
as follows. Section~\ref{sec:Model} describes the DLRSM gauge, scalar,
and fermion sectors and establishes our notation. Section~\ref{sec:ISS}
presents the ISS mass matrix, its diagonalization, and the Casas-{}-Ibarra
parametrization used in our numerical scan. Section~\ref{sec:Oneloop_g2}
derives the one-loop contributions to $a_{\mu}$ from each sector
analytically. Section~\ref{sec:Numerical-Analysis} presents the
numerical analysis and the resulting constraints on the DLRSM parameter
space. Section~\ref{sec:Conclusions} summarizes our conclusions.
The neutral gauge boson diagonalization and the complete set of master
loop-integral formulas are collected in Appendices~\ref{sec:Gauge_diagonalization}
and~\ref{sec:amu_formulae}, respectively.

\section{The Doublet Left-Right Symmetric Model}\label{sec:Model}

The DLRSM realizes parity restoration at high energies by promoting
the right-handed current to a full $SU\left(2\right)_{R}$ gauge interaction.
The gauge group is 
\begin{equation}
G_{\mathrm{LR}}=SU\left(2\right)_{L}\otimes SU\left(2\right)_{R}\otimes U\left(1\right)_{B-L},
\end{equation}
augmented by a discrete parity symmetry $\mathcal{P}$ that interchanges
the L and R sectors \cite{SENJANOVIC1979334}. The model reduces to
the SM after two successive stages of spontaneous symmetry breaking
(SSB):
\begin{equation}
G_{\mathrm{LR}}\to SU(2)_{L}\otimes U(1)_{Y}\to U(1)_{\mathrm{em}}.
\end{equation}
The first breaking, driven by the VEV $v_{R}$ of the right-handed
doublet $\chi_{R}$, generates masses for the heavy gauge bosons $W^{\prime}$
and $Z^{\prime}$ and for the right-handed neutrinos $\nu_{R}$. The
second breaking, driven by the bidoublet VEV $k_{1}$, produces the
SM fermion masses in the standard way.

\subsection{Fermion content and charge assignments}

Quarks and leptons are assigned to LR-symmetric doublet representations.
Denoting the generation index by $i=1,2,3$, the lepton and quark
doublets and their gauge quantum numbers $\left(T_{L},T_{R},B-L\right)$
are
\begin{equation}
L_{iL}=\begin{pmatrix}\nu_{i}^{\prime}\\
\ell_{i}^{\prime}
\end{pmatrix}_{L}:\left(2,1,-1\right),\qquad L_{iR}=\begin{pmatrix}\nu_{i}^{\prime}\\
\ell_{i}^{\prime}
\end{pmatrix}_{R}:\left(1,2,-1\right)
\end{equation}
\begin{equation}
Q_{iL}=\begin{pmatrix}u_{i}^{\prime}\\
d_{i}^{\prime}
\end{pmatrix}_{L}:\left(2,1,1/3\right),\qquad Q_{iR}=\begin{pmatrix}u_{i}^{\prime}\\
d_{i}^{\prime}
\end{pmatrix}_{R}:\left(1,2,1/3\right).
\end{equation}
Under parity, the left and right fermion doublets interchange $\Psi_{L}\leftrightarrow\Psi_{R}$,
for $\Psi=Q,L$. 

The electric charge is given by the generalized Gell-Mann-Nishijima
relation
\[
Q=T_{3L}+T_{3R}+\frac{B-L}{2}.
\]
To accommodate neutrino masses via the inverse seesaw mechanism (Section
\ref{sec:ISS}), we supplement the fermion content with three gauge-singlet
fermions $S_{i}$ ($i=1,2,3$) which transform under parity as
\begin{equation}
S_{i}\longleftrightarrow S_{i}^{c}.\label{eq:LR_tranformstion_fermions}
\end{equation}

We adopt the Manifest Left-Right Symmetry (MLRS) condition $g_{L}=g_{R}\equiv g$,
so the covariant derivatives acting on the fermion doublets are
\begin{equation}
D_{\mu}\Psi_{L}=\left(\partial_{\mu}-ig\frac{\vec{\tau}}{2}\cdot\vec{W}_{L\mu}-ig_{BL}\frac{\left(B-L\right)}{2}B_{\mu}\right)\Psi_{L},
\end{equation}
\begin{equation}
D_{\mu}\Psi_{R}=\left(\partial_{\mu}-ig\frac{\vec{\tau}}{2}\cdot\vec{W}_{R\mu}-ig_{BL}\frac{\left(B-L\right)}{2}B_{\mu}\right)\Psi_{R}.
\end{equation}
After the second-stage breaking $SU(2)_{R}\otimes U(1)_{B-L}\to U(1)_{Y}$,
the SM hypercharge is identified as $Y=T_{3R}+\left(B-L\right)/2$.

\subsection{Scalar sector}\label{subsec:scalar-sector}

The scalar sector consists of one bidoublet and two doublets,
\begin{equation}
\Phi\left(2,2^{*},0\right),\begin{array}{ll}
\chi_{L}\left(2,1,1\right), & \chi_{R}\left(1,2,1\right)\end{array}
\end{equation}
with the parity assignments 
\begin{equation}
\chi_{L}\longleftrightarrow\chi_{R},\qquad\Phi\longleftrightarrow\Phi^{\dagger}.\label{eq:LR_transformation_multiplets}
\end{equation}
 The field content is
\begin{equation}
\Phi=\begin{pmatrix}\phi_{1}^{0} & \phi_{1}^{+}\\
\phi_{2}^{-} & \phi_{2}^{0}
\end{pmatrix};\quad\tilde{\Phi}=\sigma_{2}\Phi^{*}\sigma_{2},\quad\chi_{L,R}=\begin{pmatrix}\chi_{L,R}^{+}\\
\chi_{L,R}^{0}
\end{pmatrix}_{L,R},
\end{equation}
and their VEVs are 
\begin{align}
\left\langle \Phi\right\rangle =\frac{1}{\sqrt{2}}\begin{pmatrix}k_{1} & 0\\
0 & k_{2}
\end{pmatrix},\quad\left\langle \chi_{R}\right\rangle =\frac{1}{\sqrt{2}}\begin{pmatrix}0\\
v_{R}
\end{pmatrix},\quad\left\langle \chi_{L}\right\rangle =\frac{1}{\sqrt{2}}\begin{pmatrix}0\\
v_{L}
\end{pmatrix}.\label{eq:vev_multipletes}
\end{align}
Hereafter we set $k_{2}=v_{L}=0$, which eliminates $W-W^{\prime}$mixing
(see Section \ref{subsec:Kinetic_gauge_sector}) and simplifies the
scalar mass spectrum.

The scalar potential consistent with the gauge symmetry and parity
is\cite{SENJANOVIC1979334}
\begin{equation}
V=V_{\Phi}+V_{\chi}+V_{\Phi\chi},\label{eq:Scalar_potential}
\end{equation}
with
\begin{align}
V_{\Phi}= & -\mu_{1}^{2}\operatorname{Tr}\Phi^{\dagger}\Phi+\lambda_{1}\left(\operatorname{Tr}\Phi^{\dagger}\Phi\right)^{2}\nonumber \\
 & +\lambda_{2}\operatorname{Tr}\Phi^{\dagger}\Phi\Phi^{\dagger}\Phi\nonumber \\
 & +\frac{1}{2}\lambda_{3}\left(\operatorname{Tr}\Phi^{\dagger}\tilde{\Phi}+\operatorname{Tr}\tilde{\Phi}^{\dagger}\Phi\right)^{2}\nonumber \\
 & +\frac{1}{2}\lambda_{4}\left(\operatorname{Tr}\Phi^{\dagger}\tilde{\Phi}-\operatorname{Tr}\tilde{\Phi}^{\dagger}\Phi\right)^{2}\nonumber \\
 & +\lambda_{5}\operatorname{Tr}\Phi^{\dagger}\Phi\tilde{\Phi}^{\dagger}\tilde{\Phi}\nonumber \\
 & +\frac{1}{2}\lambda_{6}\left[\operatorname{Tr}\Phi^{\dagger}\tilde{\Phi}\Phi^{\dagger}\tilde{\Phi}+\text{ h.c. }\right],
\end{align}
\begin{align}
V_{\chi}= & -\mu_{2}^{2}\left(\chi_{\mathrm{L}}^{\dagger}\chi_{\mathrm{L}}+\chi_{\mathrm{R}}^{\dagger}\chi_{\mathrm{R}}\right)\nonumber \\
 & +\rho_{1}\left(\left(\chi_{\mathrm{L}}^{\dagger}\chi_{\mathrm{L}}\right)^{2}+\left(\chi_{\mathrm{R}}^{\dagger}\chi_{\mathrm{R}}\right)^{2}\right)\nonumber \\
 & +\rho_{2}\chi_{\mathrm{L}}^{\dagger}\chi_{\mathrm{L}}\chi_{\mathrm{R}}^{\dagger}\chi_{\mathrm{R}},
\end{align}
\begin{align}
V_{\Phi\chi}= & \alpha_{1}\operatorname{Tr}\Phi^{\dagger}\Phi\left(\chi_{\mathrm{L}}^{\dagger}\chi_{\mathrm{L}}+\chi_{\mathrm{R}}^{\dagger}\chi_{\mathrm{R}}\right)\nonumber \\
 & +\alpha_{2}\left(\chi_{\mathrm{L}}^{\dagger}\Phi\Phi^{\dagger}\chi_{\mathrm{L}}+\chi_{\mathrm{R}}^{\dagger}\Phi^{\dagger}\Phi\chi_{\mathrm{R}}\right)\nonumber \\
 & +\alpha_{3}\left(\chi_{\mathrm{L}}^{\dagger}\tilde{\Phi}\tilde{\Phi}^{\dagger}\chi_{\mathrm{L}}+\chi_{\mathrm{R}}^{\dagger}\tilde{\Phi}^{\dagger}\tilde{\Phi}\chi_{R}\right).
\end{align}

All parameters $\mu_{1,2}^{2}$, $\lambda_{1,2,3,4,5,6}$, $\rho_{1,2}$,
and $\alpha_{1,2,3}$ are taken real, so that explicit CP violation
is absent. 

\subsubsection{Physical scalar spectrum.}

The scalar sector contains 16 real degrees of freedom from the fields
$\Phi$, $\chi_{L}$ and $\chi_{R}$. After SSB, six are eaten by
$W$, $W^{\prime}$, $Z$, $Z^{\prime}$ leaving ten physical Higgs
states. In the limit of $k_{1}\ll v_{R}$ the mass eigenstates and
their approximate masses are summarised in Table \ref{tab:scalar_masses}.
We define 
\begin{align}
\alpha_{23} & =\alpha_{2}-\alpha_{3}, & \alpha_{13} & =\alpha_{1}+\alpha_{3},\nonumber \\
\lambda_{12} & =\lambda_{1}+\lambda_{2}, & \rho_{21} & =\rho_{2}-2\rho_{1},\nonumber \\
\lambda_{2356} & =-\lambda_{2}+4\lambda_{3}+\lambda_{5}+\lambda_{6}, & \lambda_{2456} & =-\lambda_{2}-4\lambda_{4}+\lambda_{5}-\lambda_{6}.\label{eq:parameter_definitions1}
\end{align}
 
\begin{table}
\begin{centering}
\begin{tabular}{|c|c|c|c|}
\hline 
State  & $J^{P}$ & Role & Approximate mass$^{2}$\tabularnewline
\hline 
\hline 
$H_{1}^{0}$ & $0^{+}$ & SM-like Higgs & $\left(2\lambda_{12}-\alpha_{13}^{2}/2\rho_{1}\right)k_{1}^{2}$\tabularnewline
\hline 
$H_{2}^{0}$ & $0^{+}$ & Heavy CP-even & $2\rho_{1}v_{R}^{2}$\tabularnewline
\hline 
$H_{3}^{0}$ & $0^{+}$ & Heavy CP-even & $\frac{1}{2}\alpha_{23}v_{R}^{2}$\tabularnewline
\hline 
$A_{1}^{0}$ & $0^{-}$ & Heavy CP-odd & $\frac{1}{2}\alpha_{23}v_{R}^{2}$\tabularnewline
\hline 
$H_{4}^{0}$ & $0^{+}$ & Inert & $\frac{1}{2}\rho_{21}v_{R}^{2}$\tabularnewline
\hline 
$A_{2}^{0}$ & $0^{-}$ & Inert & $\frac{1}{2}\rho_{21}v_{R}^{2}$\tabularnewline
\hline 
$H_{R}^{\pm}$ & $0^{\pm}$ & Heavy charged & $\frac{1}{2}\alpha_{23}v_{R}^{2}$\tabularnewline
\hline 
$H_{L}^{\pm}$ & $0^{\pm}$ & Heavy charged & $\frac{1}{2}\rho_{21}v_{R}^{2}$\tabularnewline
\hline 
\end{tabular}
\par\end{centering}
\caption{Physical Higgs boson states and their leading-order masses in the
limit $k_{1}\ll v_{R}$.}\label{tab:scalar_masses}
\end{table}

The neutral fields $\phi_{1,2}^{0}$, $\chi_{R,L}^{0}$ can be decomposed
in terms of real and imaginary part, ($\phi\to\left(\left\langle \phi\right\rangle +\text{Re}\left(\phi\right)+i\text{Im}\left(\phi\right)\right)/\sqrt{2}$
with $\phi=\phi_{1,2}^{0},\chi_{R,L}^{0}$). In the limit of $k_{1}\ll v_{R}$,
the two CP-even states mix according to
\begin{align}
\text{Re}\left(\phi_{1}^{0}\right)\approx & \ensuremath{\frac{k_{1}}{2\rho_{1}v_{R}}\alpha_{13}H_{2}^{0}+H_{1}^{0}},\nonumber \\
\text{Re}\left(\chi_{R}^{0}\right)\approx & -\frac{k_{1}}{2\rho_{1}v_{R}}\alpha_{13}H_{1}^{0}+H_{2}^{0},\label{eq:neutral_scalar_mix}
\end{align}
where $H_{1}^{0}$ is identified with the SM Higgs and $H_{2}^{0}$
is a heavy neutral Higgs. The charged physical state $H_{R}^{+}$
and the would-be Goldstone boson $G_{R}^{\pm}$ arise from the mixing
\begin{align}
\chi_{R}^{\pm}\approx & \frac{k_{1}H_{R}^{\pm}}{v_{R}}+G_{R}^{\pm},\nonumber \\
\phi_{1}^{\pm}\approx & {\displaystyle -\frac{k_{1}G_{R}^{\pm}}{v_{R}}+H_{R}^{\pm}}.\label{eq:charged_scalar_mix}
\end{align}
The remaining physical states and their field identifications are
\begin{align}
\text{Im}\left(\chi_{L}^{0}\right)= & A_{2}^{0}, & \phi_{2}^{\pm}\approx & G_{L}^{\pm},\nonumber \\
\text{Im}\left(\chi_{R}^{0}\right)= & G_{Z^{\prime}}, & \chi_{L}^{\pm}\approx & H_{L}^{\pm},\nonumber \\
\text{Im}\left(\phi_{1}^{0}\right)= & G_{Z}, & \text{Re}\left(\phi_{2}^{0}\right)\approx & H_{3}^{0},\nonumber \\
\text{Im}\left(\phi_{2}^{0}\right)= & A_{1}^{0}, & \text{Re}\left(\chi_{L}^{0}\right)\approx & H_{4}^{0}.\label{eq:scalars_physical_basis}
\end{align}
where $G_{L}^{\pm}$, $G_{Z}$ and $G_{Z^{\prime}}$ are the Goldstone
bosons eaten by $W$, $Z$ and $Z^{\prime}$, respectively.

At leading order in $v_{R}$ the states $H_{3}^{0}$, $A_{1}^{0}$
and $H_{R}^{\pm}$ are all degenerate with common mass $M^{2}\approx\frac{1}{2}\alpha_{23}v_{R}^{2}$.
Their mass splitting for $H_{3}^{0}$ and $A_{1}^{0}$ at order $k_{1}^{2}$
is controlled by
\begin{equation}
M_{H_{3}^{0}}^{2}-M_{A_{1}^{0}}^{2}=k_{1}^{2}\delta^{0},\quad\delta^{0}\equiv\lambda_{2356}-\lambda_{2456},\label{eq:delta0}
\end{equation}
and is negligible in the $k_{1}\ll v_{R}$ employed in the numerical
analysis, their contributions to $a_{\mu}$ nearly cancel each other
in the limit $\delta^{0}\to0$. Similarly, $H_{4}^{0}$ and $A_{2}^{0}$
are exactly degenerate, however, they are inert with respect to the
external lepton lines and do not contribute to $a_{\mu}$.

\subsection{Gauge sector}\label{subsec:Kinetic_gauge_sector}

The kinetic Lagrangian for the scalar multiplets,
\[
\mathcal{L}_{D}=\left(D_{\mu}\chi_{L}\right)^{\dagger}D_{\mu}\chi_{L}+\left(D_{\mu}\chi_{R}\right)^{\dagger}D_{\mu}\chi_{R}+\operatorname{Tr}\left[\left(D_{\mu}\Phi\right)^{\dagger}D_{\mu}\Phi\right],
\]
with
\begin{align*}
D_{\mu}\chi_{L}= & \partial_{\mu}\chi_{L}-\frac{i}{2}g\vec{\tau}\cdot\vec{W}_{L}\chi_{L}-\frac{i}{2}g_{B-L}B_{\mu},\\
D_{\mu}\chi_{R}= & \partial_{\mu}\chi_{R}-\frac{i}{2}g\vec{\tau}\cdot\vec{W}_{R}\chi_{R}-\frac{i}{2}g_{B-L}B_{\mu},\\
D_{\mu}\Phi= & \partial_{\mu}\Phi-\frac{i}{2}g\left(\vec{\tau}\cdot\vec{W}_{L}\Phi-\Phi\vec{\tau}\cdot\vec{W}_{R}\right),
\end{align*}
generates the gauge boson mass matrices after SSB.

Defining $W_{L,R\mu}^{\pm}\equiv\left(W_{L,R\mu}^{1}\mp iW_{L,R\mu}^{2}\right)/\sqrt{2}$,
the charged gauge boson mass matrix in the basis $\left(W_{L}^{\pm},W_{R}^{\pm}\right)$
with $k_{2}=0$ is block diagonal, 
\begin{equation}
M_{W^{\pm}}^{2}=\frac{g^{2}}{4}\begin{pmatrix}k_{1}^{2} & 0\\
0 & k_{1}^{2}+v_{R}^{2}
\end{pmatrix}.\label{eq:Matrix_Wpm}
\end{equation}
The absence of off-diagonal entries follows directly from $k_{2}=0$
(the off-diagonal element $\propto k_{1}k_{2}$), so $W$ and $W^{\prime}$
do not mix in our approximation. The physical masses are
\begin{align*}
m_{W}^{2}\approx\frac{g^{2}k_{1}^{2}}{4},\quad m_{W^{\prime}}^{2}\approx\frac{g^{2}v_{R}^{2}}{4}.
\end{align*}

In the basis $\left(W_{\mu L}^{3},W_{\mu R}^{3},B_{\mu}\right)$,
the neutral gauge boson mass matrix is
\begin{equation}
M_{Z}^{2}=\frac{1}{4}\begin{pmatrix}g^{2}k_{1}^{2} & -g^{2}k_{1}^{2} & 0\\
-g^{2}k_{1}^{2} & g^{2}\left(k_{1}^{2}+v_{R}^{2}\right) & -gg_{B-L}v_{R}^{2}\\
0 & -gg_{B-L}v_{R}^{2} & g_{B-L}^{2}v_{R}^{2}
\end{pmatrix}\label{eq:MZ_mass_matrix}
\end{equation}
which is diagonalized in two steps (see Appendix \ref{sec:Gauge_diagonalization}).
In the limit $k_{1}\ll v_{R}$ the $Z-Z^{\prime}$ mixing angle satisfies
$\zeta\approx0$, and the gauge eigenstates are expressed in terms
of the mass eigenstates ($A_{\mu}$, $Z_{\mu}$, $Z_{\mu}^{\prime}$)
as 
\begin{align}
W_{\mu L}^{3}= & \ensuremath{{\displaystyle A_{\mu}\sin\theta_{W}-Z_{\mu}\cos\theta_{W}}},\nonumber \\
W_{\mu R}^{3}= & \ensuremath{{\displaystyle A_{\mu}\sin\theta_{W}+Z_{\mu}\sin\theta_{W}\tan\theta_{W}-Z_{\mu}^{\prime}\frac{\sqrt{\cos\left(2\theta_{W}\right)}}{\cos\theta_{W}}},}\nonumber \\
B_{\mu}= & \ensuremath{{\displaystyle A_{\mu}\sqrt{\cos\left(2\theta_{W}\right)}+Z_{\mu}\sqrt{\cos\left(2\theta_{W}\right)}\tan\theta_{W}+Z_{\mu}^{\prime}\tan\theta_{W}}}.\label{eq:Neutral_guage_mixings}
\end{align}
The physical neutral gauge boson masses in this limit are 
\begin{align*}
m_{Z}^{2}=\frac{m_{W}^{2}}{\cos^{2}\theta_{W}},\quad m_{Z^{\prime}}^{2}=\ensuremath{m_{W^{\prime}}^{2}\frac{\cos^{2}\theta_{W}}{\cos\left(2\theta_{W}\right)}{\displaystyle -m_{W}^{2}\frac{\left(\tan\left(2\theta_{W}\right)+4\right)\tan^{2}\theta_{W}}{4}}.}
\end{align*}
From the covariant derivatives and the field rotation \eqref{eq:Neutral_guage_mixings},
the interactions of charged leptons with the neutral gauge bosons
are
\begin{align}
\mathcal{L}_{Z\ell\ell} & =\frac{ig}{4\cos\theta_{W}}Z\overline{\ell}\gamma^{\mu}\left[\left(3\sin^{2}\theta_{W}-\cos^{2}\theta_{W}\right)+\gamma^{5}\right]\ell,\nonumber \\
\mathcal{L}_{Z^{\prime}\ell\ell} & =\frac{ig}{4\cos\theta_{W}\sqrt{\cos\left(2\theta_{W}\right)}}Z^{\prime}\overline{\ell}\gamma^{\mu}\left[\left(2\sin^{2}\theta_{W}-\cos\left(2\theta_{W}\right)\right)-\cos\left(2\theta_{W}\right)\gamma^{5}\right]\ell.\label{eq:ZZp_ll_interactions}
\end{align}

\section{The inverse seesaw }\label{sec:ISS}

The DLRSM naturally accommodates the inverse seesaw (ISS) mechanism
as an alternative to the standard seesaw. Whereas in type-I seesaw
the righ handed neutrino mass $M_{R}\sim10^{14}$ Gev is inaccesibly
large, the ISS achieves the observed light neutrino mass scale via
a small lepton-number-breaking parameter $\mu_{S}$, keeping $M_{D}\propto v_{R}$
at the TeV scale where it can be probed at colliders.

\subsection{Yukawa Lagrangian and mass matrices}

The lepton Yukawa Lagrangian, invariant under $\mathrm{G_{LR}}$ and
the parity symmetry, is \cite{Gu:2010xc,Brdar:2018sbk} 
\begin{equation}
\begin{aligned}-\mathcal{L}_{Y}= & \overline{L}_{L}Y\Phi L_{R}+\overline{L}_{L}\tilde{Y}\tilde{\Phi}L_{R}+\overline{S}Y_{L}\tilde{\chi}_{L}^{\dagger}L_{L}+\overline{S^{c}}Y_{R}\tilde{\chi}_{R}^{\dagger}L_{R}+\frac{1}{2}\overline{S}^{c}\mu_{S}S+\text{h.c.},\end{aligned}
\label{eq:LYukawa}
\end{equation}

where $Y$, $\tilde{Y}$, $Y_{L}$, $Y_{R}$ are $3\times3$ complex
Yukawa matrices, $\mu_{S}$ is the Majorana mass matrix for fermionic
singlets $S_{i}$, and we use $\tilde{X}\equiv i\sigma_{2}X^{*}$
for scalars doublets and $S^{c}=C\overline{S}^{\top}$for the charge
conjugate of $S$.

The parity symmetry, combining \eqref{eq:LR_tranformstion_fermions}
and \eqref{eq:LR_transformation_multiplets}, imposes
\begin{equation}
Y_{L}=Y_{R},\qquad Y=Y^{\dagger},\qquad\tilde{Y}=\tilde{Y}^{\dagger},\qquad\mu_{S}=\mu_{S}^{\dagger},\label{eq:Yukawas_LR_symmetry}
\end{equation}

above the LR symmetry breaking scale. After SSB the lepton mass matrices
read
\begin{equation}
M_{\ell}=\frac{k_{1}}{\sqrt{2}}\tilde{Y},\quad m_{D}=\frac{k_{1}}{\sqrt{2}}Y,\quad M_{D}=\frac{v_{R}}{\sqrt{2}}Y_{R}.\label{eq:Dirac_Majorana}
\end{equation}
where $M_{\ell}$, is the charged-lepton mass matrix, $m_{D}$ is
the light-heavy Dirac neutrino mass matrix, and $M_{D}$ is the heavy
Majorana mass matrix for the right-handed neutrinos. The charged lepton
masses are extracted via a biunitary transformation,
\begin{equation}
\hat{M}_{\ell}=V_{L}^{\ell\dagger}M_{\ell}V_{R}^{\ell}=\text{diag}\left(m_{e},m_{\mu},m_{\tau}\right).\label{eq:Ml_biunitary}
\end{equation}

\subsection{Neutrino mass matrix and spectrum}

In the basis $n_{L}^{\prime}=\left(\nu_{L},\nu_{R}^{c},S^{c}\right)$,
$n_{R}^{\prime}=\left(\nu_{L}^{c},\nu_{R},S\right)$ the full $9\times9$
neutrino Majorana mass matrix takes the ISS block form
\begin{equation}
\mathcal{M}_{\nu}=\begin{pmatrix}0 & B^{\top}\\
B & C
\end{pmatrix},\quad B=\begin{pmatrix}m_{D}\\
0
\end{pmatrix},\quad C=\begin{pmatrix}0 & M_{D}^{\top}\\
M_{D} & \mu_{S}
\end{pmatrix}.\label{eq:Mnu_ISS}
\end{equation}
In the hierarchy $|C|\gg|B|$, the seesaw formula gives the light
neutrino mass matrix 
\begin{equation}
m_{\nu}\approx{\displaystyle m_{D}^{\top}M_{D}^{-1}\mu_{S}\left(M_{D}^{\top}\right)^{-1}m_{D}}.\label{eq:light_nu_approx}
\end{equation}
Note that $m_{\nu}\propto\mu_{S}$: in contrast to the type-I seesaw,
the light-neutrino masses can be small even for $M_{D}\sim v_{R}\sim\mathcal{O}\left(\text{TeV}\right)$,
provided $\mu_{S}\ll m_{D}$. The heavy states form pseudo-Dirac pairs
$N_{i}^{\pm}$with degenerate masses
\begin{align}
M_{i}^{-}\approx M_{i}^{+}\approx M_{Di},\label{eq:Mipm}
\end{align}
split at $\mathcal{O}\left(\mu_{S}\right)$.

\subsection{Mixing matrix and Casas-Ibarra parametrization}

Defining the active-{}-sterile mixing parameter
\begin{equation}
\xi\equiv m_{D}^{\top}M_{D}^{-1}\label{eq:xi_def}
\end{equation}
and imposing $M_{D}^{-1}\mu_{S}M_{D}^{-1}m_{D}\approx0$, the approximate
mixing matrix is \cite{Catano:2012kw,PhysRevD.86.035007}
\begin{equation}
\mathcal{U}\approx\begin{pmatrix}U_{\nu} & -\frac{i}{\sqrt{2}}\xi & \frac{1}{\sqrt{2}}\xi\\
0 & \frac{i}{\sqrt{2}}\mathbb{I} & \frac{1}{\sqrt{2}}\mathbb{I}\\
-\xi^{\top}U_{\nu} & -\frac{i}{\sqrt{2}}\mathbb{I} & \frac{1}{\sqrt{2}}\mathbb{I}
\end{pmatrix}\label{eq:U_total}
\end{equation}
where $U_{\nu}$ is the unitary matrix that diagonalizes the light
neutrino mass matrix \eqref{eq:light_nu_approx}. 

The $9\times9$ PMNS-like matrix $\mathcal{U}$ diagonalizes $\mathcal{M}_{\nu}$
and can be decomposed into three $3\times9$ block matrices,
\[
\mathcal{U}=\begin{pmatrix}U_{L}\\
U_{R}\\
U_{S}
\end{pmatrix},
\]
such that 
\[
\hat{\mathcal{M_{\nu}}}=\mathcal{U}^{\top}\mathcal{M}_{\nu}\mathcal{U}=\text{diag}\left(m_{i},M_{i}^{-},M_{i}^{+}\right).
\]
The unitary condition $U_{X}U_{Y}^{\dagger}=\delta_{XY}\mathbb{I}$
$\left(X,Y=L,R,S\right)$ yields the identities
\begin{equation}
m_{D}=U_{R}^{*}\hat{\mathcal{M_{\nu}}}U_{L}^{\dagger},\quad M_{D}=U_{S}^{*}\hat{\mathcal{M_{\nu}}}U_{R}^{\dagger},\quad\mu_{S}=U_{S}^{*}\hat{\mathcal{M}_{\nu}}U_{S}^{\dagger}.\label{eq:Mdiag_mDMDmu}
\end{equation}

The light-neutrino \eqref{eq:light_nu_approx} can be rewritten as
\begin{equation}
m_{\nu}\approx m_{D}^{\top}\mathcal{M}^{-1}m_{D};\qquad\mathcal{M}=M_{D}\mu_{S}^{-1}M_{D}^{\top}.\label{eq:ISS_approxM}
\end{equation}
and by means of the Casas-Ibarra parametrization the Dirac neutrino
mass matrix is written as follows \cite{Casas:2001sr,PhysRevD.91.015001}
\begin{equation}
\begin{aligned}m_{D}=V^{\dagger}\text{diag}\left(\sqrt{\mathcal{M}_{1}},\sqrt{\mathcal{M}_{2}},\sqrt{\mathcal{M}_{3}}\right)R\text{diag}\left(\sqrt{m_{\nu_{1}}},\sqrt{m_{\nu_{2}}},\sqrt{m_{\nu_{3}}}\right)U_{\nu}^{\dagger},\end{aligned}
\label{eq:Casas-Ibarra1}
\end{equation}
where $V$ diagonalises $\mathcal{M}$ and $R$ is a complex orthogonal
matrix $\left(RR^{\top}=\mathbb{I}\right)$ that parametrises the
remaining freedom in $m_{D}$ consistent with the light neutrino spectrum.
In the simplified diagonal ansatz used in the numerical analysis,
$M_{D}=\text{diag}\left(M_{D1},M_{D2},M_{D3}\right)$, $\mu_{S}=\text{diag}\left(\mu_{S1},\mu_{S2},\mu_{S3}\right)$,
$R=\mathbb{I}$, then, $V=\mathbb{I}$ and

\begin{equation}
\begin{aligned}m_{D}= & \text{diag}\left(\sqrt{\tfrac{m_{\nu_{1}}}{\mu_{S1}}}M_{D1},\;\sqrt{\tfrac{m_{\nu_{2}}}{\mu_{S2}}}M_{D2},\;\sqrt{\tfrac{m_{\nu_{3}}}{\mu_{S3}}}M_{D3}\right)U_{\nu}^{\dagger}\end{aligned}
\label{eq:Casas-Ibarra2}
\end{equation}

\subsection{Neutrino mass basis and charged currents}

The weak neutrino states are rotated to the mass eigenstates $n=\left(\nu_{i},N_{i}^{-},N_{i}^{+}\right)$
by
\begin{align*}
n_{L}^{\prime}=\mathcal{U}^{*}n_{L},\quad n_{R}^{\prime}=\mathcal{U}n_{R}.
\end{align*}
The charged-current Lagrangian mediated by $W$, $W^{\prime}$is
\begin{equation}
\mathcal{L}_{W}=\frac{g}{2}W^{+}\overline{n_{i}}\gamma^{\mu}P_{L}Q_{L,ai}^{*}\ell_{a}+\frac{g}{2}W^{\prime+}\overline{n_{i}}\gamma^{\mu}P_{R}Q_{R,ai}^{*}\ell_{a}+\text{h.c.},\label{eq:WWplninteractions}
\end{equation}
where the left and right leptonic mixing matrices are
\begin{align}
Q_{L}= & V_{L}^{\ell\dagger}U_{L}^{*}, & Q_{R}= & V_{R}^{\ell\dagger}U_{R}.\label{eq:QLR_def}
\end{align}

\subsection{Yukawa interactions}

In the unitary gauge the Goldstone bosons are absent; retaining only
physical scalars (see Section \ref{subsec:scalar-sector}) the Yukawa
Lagrangian \eqref{eq:LYukawa} decomposes into neutral and charged
parts. 

The interactions of neutral Higgs bosons with charged leptons read
\begin{align}
-\mathcal{L}_{Y}^{0}\supset & \frac{m_{\ell}}{k_{1}}h^{SM}\overline{\ell}\ell+\frac{m_{\ell}\alpha_{13}}{2\rho_{1}v_{R}}H_{2}^{0}\overline{\ell}\ell+\left(\frac{1}{k_{1}}H_{3}^{0}\overline{\ell}\Lambda P_{R}\ell+\frac{i}{k_{1}}A_{1}^{0}\overline{\ell}\Lambda P_{R}\ell+\text{h.c.}\right)\label{eq:LY_neutralHiggs}
\end{align}
where
\begin{align}
\Lambda= & V_{L}^{\ell\dagger}m_{D}V_{R}^{\ell}.\label{eq:Lambda}
\end{align}
The matrix $\Lambda$ in \eqref{eq:Lambda} is the combination of
$m_{D}$ and the charged leptonic diagonalisation matrices that controls
the flavour-changing couplings of $H_{3}^{0}$ and $A_{1}^{0}$ to
charged leptons. Note that both $H_{3}^{0}$ and $A_{1}^{0}$ enter
through the same combination $\Lambda P_{R}$; the only difference
between their couplings is the factor of $i$ for $A_{1}^{0}$ in
\eqref{eq:LY_neutralHiggs}, which is a pure phase and does not affect
$\left|\Lambda\right|^{2}$ in the squared amplitude, the reason both
contribute to $a_{\mu}$ with the same parametric dependence (see
Section \ref{subsec:contribution_Sll}).

The charged scalars interactions with leptons and neutrinos are
\begin{align}
-\mathcal{L}_{Y}^{\pm}=\frac{1}{\sqrt{2}}H_{R}^{+}\overline{n}\left(\left(K-\frac{k_{1}}{v_{R}}D\right)P_{R}-\tilde{T}P_{L}\right)\ell-\frac{1}{\sqrt{2}}H_{L}^{+}\overline{n}EP_{L}\ell+\text{h.c.}\label{eq:LY_chargedhiggs}
\end{align}
with the mixing matrices defined as
\begin{align}
K= & U_{L}^{\top}YV_{R}^{\ell} & D= & U_{S}^{\top}Y_{R}V_{R}^{\ell}\nonumber \\
\tilde{T}= & U_{R}^{\dagger}\tilde{Y}^{\dagger}V_{L}^{\ell}, & E= & U_{S}^{\dagger}Y_{L}V_{L}^{\ell}.\label{eq:Mixing_matrices}
\end{align}

\section{One loop contributions to $\left(g-2\right)_{\mu}$}\label{sec:Oneloop_g2}

Within the DLRSM, supplemented by the inverse seesaw, the anomalous
magnetic moment of the muon receives one-loop corrections from five
classes of new particles: the neutral gauge boson $Z^{\prime}$, the
charged gauge boson $W^{\prime}$, the heavy charged scalars $H_{L,R}^{\pm}$
and the neutral scalars $H_{2,3}^{0}$, $A_{1}^{0}$. Table \ref{tab:amu_diagrams}
lists the diagram topologies and their internal particle content.
Troughout this section we work in the unitary gauge, set $V_{L}^{\ell}=V_{R}^{\ell}=\mathbb{I}$
for simplicity, and use the general formulae for $a_{\mu}$ new physics
contributions from \cite{JEGERLEHNER20091} and summarized in Appendix
\ref{sec:amu_formulae}. The Feynman diagrams are shown in Figs. \ref{fig:Feynman_XFF}
and \ref{fig:Feynman_FXX}. Finally the total contribution to $a_{\mu}$
is given by
\begin{equation}
\Delta a_{\mu}^{\text{\text{DLRSM}}}=\sum_{\Theta}a_{\mu}\left(\Theta\right).\label{eq:amu_DLRSM_total}
\end{equation}
 
\begin{table}
\begin{centering}
\begin{tabular}{|c|c|c|c||c|c|c|c|}
\hline 
$\Theta$ & $P_{0}$ & $P_{1}$ & $P_{2}$ & $\Theta$ & $P_{0}$ & $P_{1}$ & $P_{2}$\tabularnewline
\hline 
\hline 
FVV & $n_{i}$ & $W^{\prime}$ & $W^{\prime}$ & VFF & $Z^{\prime}$ & $\mu$ & $\mu$\tabularnewline
\hline 
FSS & $n_{i}$ & $H_{R}^{\pm}$ & $H_{R}^{\mp}$ & SFF & $H_{i}^{0}$ & $\ell$ & $\ell$\tabularnewline
\hline 
FSS & $n_{i}$ & $H_{L}^{\pm}$ & $H_{L}^{\mp}$ & SFF & $A_{1}^{0}$ & $\ell$ & $\ell$\tabularnewline
\hline 
\end{tabular}
\par\end{centering}
\caption{One loop diagram topologies contributing to $a_{\mu}$ in the DLRSM.
The notation $\Theta\left(P_{0},P_{1},P_{2}\right)$ identifies the
topology and the particles in the loop: $P_{0}$ is the internal neutral
fermion (F) or boson (B) connecting to the external muon lines, while
$P_{1}$ and $P_{2}$ are the two internal propagators closing the
loop.}\label{tab:amu_diagrams}
\end{table}

\subsection{Neutral contributions SFF and VFF topologies}\label{subsec:contribution_Sll}

\begin{figure}
\begin{centering}
\includegraphics[scale=0.4]{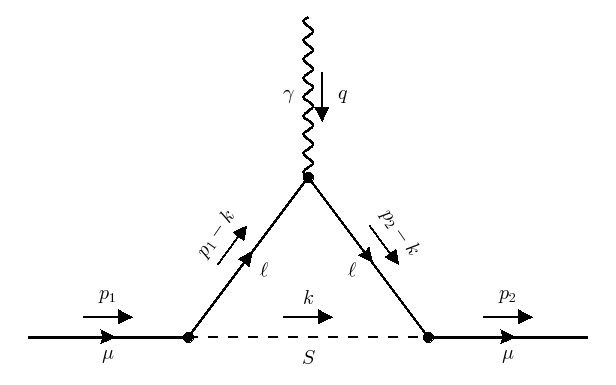}\includegraphics[scale=0.4]{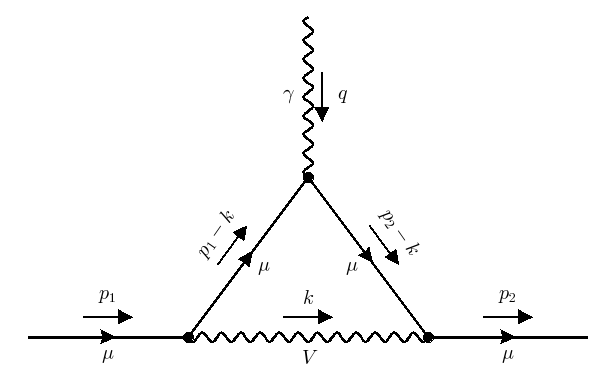}
\par\end{centering}
\caption{Feynman diagrams for neutral exchange bosons SFF (left) and VFF (right)
topologies}\label{fig:Feynman_XFF}

\end{figure}

For the neutral scalars, $a_{\mu}$ has contributions of the left
Feynman diagram in Figure \ref{fig:Feynman_XFF}, we consider the
approximation $m_{\ell}\ll M_{S}$ ($\lambda\to0$) with $S=H_{2,3}^{0},A_{1}^{0}$.
From \eqref{eq:LY_neutralHiggs}, the couplings of $H_{2,3}^{0}$
and $A_{1}^{0}$ with contributions to $a_{\mu}$ are, respectively
\begin{align}
H_{2}^{0}: & C_{S}=\frac{m_{\mu}\alpha_{13}}{2\rho_{1}v_{R}}, & C_{P}=0;\\
H_{3}^{0}: & C_{S}=-\frac{1}{\sqrt{2}}Y_{\ell\mu}, & C_{P}=0;\\
A_{1}^{0}: & C_{S}=0, & C_{P}=-\frac{i}{\sqrt{2}}Y_{\ell\mu},
\end{align}
for $Y$ real \eqref{eq:Dirac_Majorana}, so that,
\begin{align}
a_{\mu}\left(H_{2}^{0}\mu\mu\right)\approx & \frac{m_{\mu}^{2}}{16\pi^{2}M_{H_{2}^{0}}^{2}}\left(\frac{m_{\mu}\alpha_{13}}{2\rho_{1}v_{R}}\right)^{2}\left(4\log\left(\frac{m_{\mu}}{M_{H_{2}^{0}}}\right)+3\right),\nonumber \\
a_{\mu}\left(H_{3}^{0}\ell\ell\right)\approx & \frac{1}{32\pi^{2}}\frac{m_{\mu}}{M_{H_{3}^{0}}^{2}}\sum_{\ell}m_{\ell}\left|Y_{\ell\mu}\right|^{2}\left(4\log\left(\frac{m_{\ell}}{M_{H_{3}^{0}}}\right)+3\right),\nonumber \\
a_{\mu}\left(A_{1}^{0}\ell\ell\right)\approx & -\frac{1}{32\pi^{2}}\frac{m_{\mu}}{M_{A_{1}^{0}}^{2}}\sum_{\ell}m_{\ell}\left|Y_{\ell\mu}\right|^{2}\left(4\log\left(\frac{m_{\ell}}{M_{A_{1}^{0}}}\right)+3\right).\label{eq:amu_Hi0}
\end{align}
 The contributions of neutral scalars $S=H_{2,3}^{0},A_{1}^{0}$
are proportional to $\left(\frac{m_{\mu}}{M_{S}}\right)^{2}$, but
in particular the $H_{2}^{0}$ contribution is suppressed additionally
by the coupling of interaction $H_{2}^{0}\mu\mu$ which give a factor
proportional to $\left(\frac{m_{\mu}}{v_{R}}\right)^{2}$, in consequence
$a_{\mu}\left(H_{2}^{0}\right)$ is negligible for large $v_{R}$.
The contributions to $a_{\mu}$ from $H_{3}^{0}$ and $A_{1}^{0}$
are of equal magnitude in the case of $\delta^{0}=0$, from \eqref{eq:delta0}
where the two masses are degenerated and cancel each other. In the
case of $\delta^{0}\neq0$, the values of $\lambda_{2356}$ and $\lambda_{2456}$
contributes negligibly at the limit $k_{1}\ll v_{R}$. Focusing on
$H_{3}^{0}$ contribution we observe that $\log\left(\frac{m_{\ell}}{M_{H_{3}^{0}}}\right)$
is expected to be negative for large values of $M_{H_{3}^{0}}$ which
is proportional to $v_{R}$, in consequence, $a_{\mu}\left(H_{3}^{0}\ell\ell\right)<0$,
the same argument applies to $H_{2}^{0}$ contribution, and for $A_{1}^{0}$
this implies $a_{\mu}\left(A_{1}^{0}\ell\ell\right)>0$.

For a neutral vector boson $V$, $a_{\mu}$ has contributions from
Feynman diagram in the right Figure \ref{fig:Feynman_XFF}. Reading
off $C_{V}$ and $C_{A}$ from the interaction Lagrangian \eqref{eq:ZZp_ll_interactions}
and substituting in \eqref{eq:amu_general} in the approximation $m_{\ell}\ll M_{Z^{\prime}}$,
\begin{align*}
a_{\mu}\left(Z^{\prime}\mu\mu\right) & \approx-\frac{1}{12\pi^{2}}\frac{g^{2}}{16\cos^{2}\theta_{W}\cos\left(2\theta_{W}\right)}\left(\frac{m_{\mu}}{M_{Z^{\prime}}}\right)^{2}\left(5\cos^{2}\left(2\theta_{W}\right)-\left(2\sin^{2}\theta_{W}-\cos\left(2\theta_{W}\right)\right)^{2}\right).
\end{align*}
With $\sin^{2}\theta_{W}\approx0.231$ the bracket evaluates to approximately
$1.5$, so $a_{\mu}\left(Z^{\prime}\right)<0$.

\subsection{Charged contributions: SFF and FVV topologies}

\begin{figure}
\begin{centering}
\includegraphics[scale=0.4]{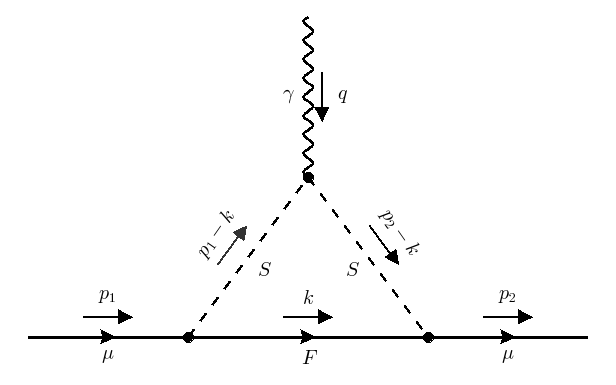}\includegraphics[scale=0.4]{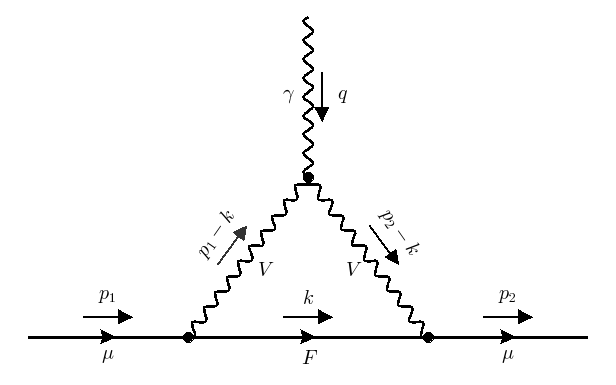}
\par\end{centering}
\caption{Feynman diagrams for the FSS (left) and FVV (right) topologies}\label{fig:Feynman_FXX}
\end{figure}

The FSS diagram have neutrinos in the loop and charged scalars $H_{L,R}^{\pm}$
as the two internal propagators (left Feynman diagram in Figure \ref{fig:Feynman_FXX}).
We treat the two neutrino mass scales separately: light neutrinos
($m_{n}\approx0$, $m_{\mu}\ll M_{S}$) and heavy neutrinos $M_{F}=m_{n}$,
($m_{\mu}\ll M_{S},\,m_{n}$). The general FSS formula in each limit
is
\begin{equation}
a_{\mu}\left(n,S^{\pm}\right)\approx\begin{cases}
-\frac{1}{48\pi^{2}}\left(\frac{m_{\mu}}{M_{S^{\pm}}}\right)^{2}\left(\left|C_{P}\right|^{2}+\left|C_{S}\right|^{2}\right) & m_{n}\approx0,m_{\mu}\ll M_{S},\\
\frac{1}{16\pi^{2}}\sum_{n=4}^{9}\left(\frac{m_{\mu}}{m_{n}}\right)\left(\left|C_{P}\right|{}^{2}-\left|C_{S}\right|{}^{2}\right)\mathcal{G}\left(\frac{M_{S^{\pm}}}{m_{n}}\right) & m_{\mu}\ll M_{S},m_{n}.
\end{cases}\label{eq:FSS_expression}
\end{equation}
where 
\begin{equation}
\mathcal{G}\left(r\right)=\frac{1-4r^{2}\log r-r^{4}}{\left(1-r^{2}\right)^{3}}.
\end{equation}

Substituting the couplings from \eqref{eq:LY_chargedhiggs} in \eqref{eq:FSS_expression}
gives the explicit $H_{L,R}^{\pm}$ contributions:
\begin{align*}
a_{\mu}\left(n,H_{L}^{\pm}\right)\approx & 0,\\
a_{\mu}\left(n,H_{R}^{\pm}\right)\approx & \frac{1}{16\pi^{2}}\frac{1}{4}\sum_{n=4}^{9}\left(\frac{m_{\mu}}{m_{n}}\right)\mathcal{G}\left(\frac{M_{H_{R}^{\pm}}}{m_{n}}\right)\text{Re}\left(K_{n\mu}\tilde{T}_{n\mu}^{*}\right),
\end{align*}
neglecting light neutrino contributions. From \eqref{eq:LY_chargedhiggs},
the interaction of $H_{L}^{\pm}$ with leptons is purely chiral in
consequence $\left|C_{P}\left(H_{L}^{\pm}\right)\right|{}^{2}-\left|C_{S}\left(H_{L}^{\pm}\right)\right|{}^{2}=0$.
The contribution $a_{\mu}\left(n,H_{R}^{\pm}\right)$ depends on $M_{H_{R}^{\pm}}/m_{n}$,
encoding the mass dependence on $\alpha_{23}$. 

For the FVV topology for charged gauge boson (right feynman diagram
in Figure \ref{fig:Feynman_FXX}) the $a_{\mu}$ contribution is given
by
\begin{equation}
a_{\mu}\left(FVV\right)\approx\begin{cases}
\frac{5}{24\pi^{2}}\left(\frac{m_{\mu}}{M_{V}}\right)^{2}\left(\left|C_{V}\right|^{2}+\left|C_{A}\right|^{2}\right) & m_{F}\approx0,m_{\mu}\ll M_{V},\\
\frac{1}{16\pi^{2}}\left[\left(\frac{m_{\mu}}{m_{n}}\right)^{2}\mathcal{F}_{+}\left(\frac{M_{V}}{m_{n}}\right)\left(\left|C_{V}\right|^{2}+\left|C_{A}\right|^{2}\right)-\left(\frac{m_{\mu}}{m_{n}}\right)\mathcal{F}_{-}\left(\frac{M_{V}}{m_{n}}\right)\left(\left|C_{V}\right|^{2}-\left|C_{A}\right|^{2}\right)\right] & m_{\mu}\ll M_{F,}M_{V}
\end{cases}\label{eq:amu_FVV}
\end{equation}
where
\begin{align*}
\mathcal{F}_{+}\left(r\right)= & \frac{1}{3r^{2}\left(1-r^{2}\right)^{3}}\left[4+35r^{2}-43r^{4}+10r^{6}\right],\\
\mathcal{F}_{-}\left(r\right)= & \frac{1}{r^{2}\left(1-r^{2}\right)^{3}}\left[1+12r^{2}+20r^{2}\log r-17r^{4}+4r^{6}\right],
\end{align*}
For the $W^{\prime}$ loop the coupling to $n_{i}\mu$ is purely chiral
\eqref{eq:WWplninteractions}, so $C_{V}=C_{A}=g\,Q_{R,n\mu}/2$.
Substituting from \eqref{eq:amu_FVV},
\begin{equation}
a_{\mu}\left(nW^{\prime}W^{\prime}\right)\approx\frac{1}{8\pi^{2}}\frac{g^{2}}{4}\left[\frac{5}{3}\left(\frac{m_{\mu}}{M_{W^{\prime}}}\right)^{2}\sum_{n=1}^{3}\left|Q_{R,n\mu}\right|^{2}+\frac{1}{2}\left(\frac{m_{\mu}}{m_{n}}\right)^{2}\sum_{n=4}^{9}\left|Q_{R,n\mu}\right|^{2}\mathcal{F}_{+}\left(\frac{M_{W^{\prime}}}{m_{n}}\right)\right].
\end{equation}

Finally, Table \ref{tab:power_counting} collects the parametric scaling
of each contribution

\begin{table}
\begin{centering}
\begin{tabular}{|c|c|c|}
\hline 
Contribution & Scaling & Sign\tabularnewline
\hline 
\hline 
$a_{\mu}\left(Z^{\prime}\right)$ & $g^{2}/v_{R}^{2}$ & $-$\tabularnewline
\hline 
$a_{\mu}\left(W^{\prime}\right)$ & $g^{2}/v_{R}^{2}$ & $+$\tabularnewline
\hline 
$a_{\mu}\left(H_{3}^{0}\right)$ & $Y_{R}^{2}/\left(\alpha_{23}v_{R}^{2}\right)$ & $-$\tabularnewline
\hline 
$a_{\mu}\left(A_{1}^{0}\right)$ & $Y_{R}^{2}/\left(\alpha_{23}v_{R}^{2}\right)$ & $+$\tabularnewline
\hline 
$a_{\mu}\left(H_{R}^{\pm}\right)$ & $Y_{R}^{2}/\left(\alpha_{23}^{2}v_{R}^{4}\right)$ & mixed\tabularnewline
\hline 
$a_{\mu}\left(H_{L}^{\pm}\right)$ & $Y_{L}^{2}/v_{R}^{4}$ & $0$\tabularnewline
\hline 
$a_{\mu}\left(H_{2}^{0}\right)$ & $\alpha_{13}^{2}m_{\mu}^{2}/v_{R}^{4}$ & $-$\tabularnewline
\hline 
\end{tabular}
\par\end{centering}
\caption{Leading parametric scaling of each one-loop contribution to $a_{\mu}$
with the DLRSM parameters.}\label{tab:power_counting}

\end{table}

\section{Numerical Analysis}\label{sec:Numerical-Analysis}

The light neutrino mixing angles and mass-squared differences are
fixed to the Normal Ordering (NO) best fit values from the NuFit collaboration
\cite{Esteban:2024eli}, summarized in Table \ref{tab:Nu_data}. These
enter the analysis through the Casas-Ibarra parametrization \eqref{eq:Casas-Ibarra2},
which expresses $m_{D}$ in terms of $M_{D}$, $\mu_{X}$ and neutrino
masses $m_{n}$. 
\begin{table}
\begin{centering}
\begin{tabular}{|c|c|c|}
\hline 
 & \multicolumn{2}{c|}{Normal Ordering (best fit)}\tabularnewline
\hline 
\hline 
 & $\text{bfp}\pm1\sigma$ & $3\sigma$ range\tabularnewline
\hline 
$\ensuremath{\sin^{2}\theta_{12}}$ & $\ensuremath{0.308_{-0.011}^{+0.012}}$ & $\ensuremath{0.275\rightarrow0.345}$\tabularnewline
\hline 
$\ensuremath{\sin^{2}\theta_{23}}$ & $\ensuremath{0.470_{-0.013}^{+0.017}}$ & $\ensuremath{0.435\rightarrow0.585}$\tabularnewline
\hline 
$\ensuremath{\sin^{2}\theta_{13}}$ & $\ensuremath{0.02215_{-0.00058}^{+0.00056}}$ & $\ensuremath{0.02030\rightarrow0.02388}$\tabularnewline
\hline 
$\ensuremath{\delta_{\mathrm{CP}}/{}^{\circ}}$ & $\ensuremath{212_{-41}^{+26}}$ & $\ensuremath{124\rightarrow364}$\tabularnewline
\hline 
$\Delta m_{21}^{2}/10^{-5}\mathrm{eV}^{2}$ & $7.49_{-0.19}^{+0.19}$ & $\ensuremath{6.92\rightarrow8.05}$\tabularnewline
\hline 
$\Delta m_{31}^{2}/10^{-3}\mathrm{eV}^{2}$ & $\ensuremath{+2.513_{-0.019}^{+0.021}}$ & $\ensuremath{+2.451\rightarrow+2.578}$\tabularnewline
\hline 
\end{tabular}
\par\end{centering}
\caption{Normal ordering best-fit values and $3\sigma$ ranges for neutrino
oscillation parameters, taken from NuFit global analysis \cite{Esteban:2024eli}.}\label{tab:Nu_data}
\end{table}

We set $\lambda_{12}=\rho_{1}=1$, which fixes the SM-like Higgs mass
condition from Table \ref{tab:scalar_masses} to 
\[
\alpha_{13}^{2}\approx2\left(2-\frac{m_{H_{1}^{0}}^{2}}{k_{1}^{2}}\right)\approx1.87,\quad M_{H_{2}^{0}}^{2}\approx\ensuremath{2\left(v_{R}^{2}+2k_{1}^{2}-m_{H_{1}^{0}}^{2}\right)\approx2v_{R}^{2}.}
\]
The masses $H_{3}^{0}$ , $A_{1}^{0}$, $H_{R}^{\pm}$ are then controlled
by $\alpha_{23}$ alone (at leading order in $v_{R}$, see Table \ref{tab:scalar_masses})
. Requiring $M_{H_{3}^{0}}>M_{H_{1}^{0}}$ sets a lower bound 
\begin{equation}
\alpha_{23}>2\frac{M_{H_{1}^{0}}^{2}}{v_{R}^{2}}.
\end{equation}
which over the range $v_{R}\in[10^{3},10^{6}]$ GeV, implies $\alpha_{23}^{\text{min}}\in[10^{-8},10^{-2}]$. 

For simplicity, we take the degenerate limit 
\begin{align}
M_{i}^{-}\approx M_{i}^{+}=M=\frac{Y_{R}}{\sqrt{2}}v_{R},\quad\mu_{S1}=\mu_{S2}=\mu_{S3}\equiv\mu_{X},\label{eq:Mipm_degenerate}
\end{align}
 which reduces the heavy neutrino space. In this limit, we could approximate
\begin{equation}
Y_{ij}\approx\frac{\sqrt{m_{\text{light}}}}{\sqrt{\mu_{X}}}\frac{v_{R}}{k_{1}}\,\mathcal{O}\left(0.1\right)\label{eq:Yij_approx}
\end{equation}
where we fix the elements of $U_{\nu}$ to be of order $\mathcal{O}\left(0.1\right)$
and $Y_{R}\sim1$, but the Yukawa couplings are constrained by perturvativity
$|Y_{ij}|^{2}<6\pi$, consequently, from \eqref{eq:Yij_approx}
\begin{equation}
\left(\frac{\text{GeV}}{\mu_{X}}\right)\left(\frac{v_{R}^{2}}{\text{GeV}^{2}}\right)\lesssim10^{19},\label{eq:Y_perturvatividad}
\end{equation}
assuming a light-neutrino mass scale of order $\mathcal{O}\left(10^{-12}\text{ GeV}\right)$.
The free parameters of the analysis are therefore $v_{R}$, $Y_{R}$,
$\mu_{X}$ and $\alpha_{23}$. 

\subsection{Dependence on individual parameters}

\begin{figure}
\begin{centering}
\includegraphics[scale=0.5]{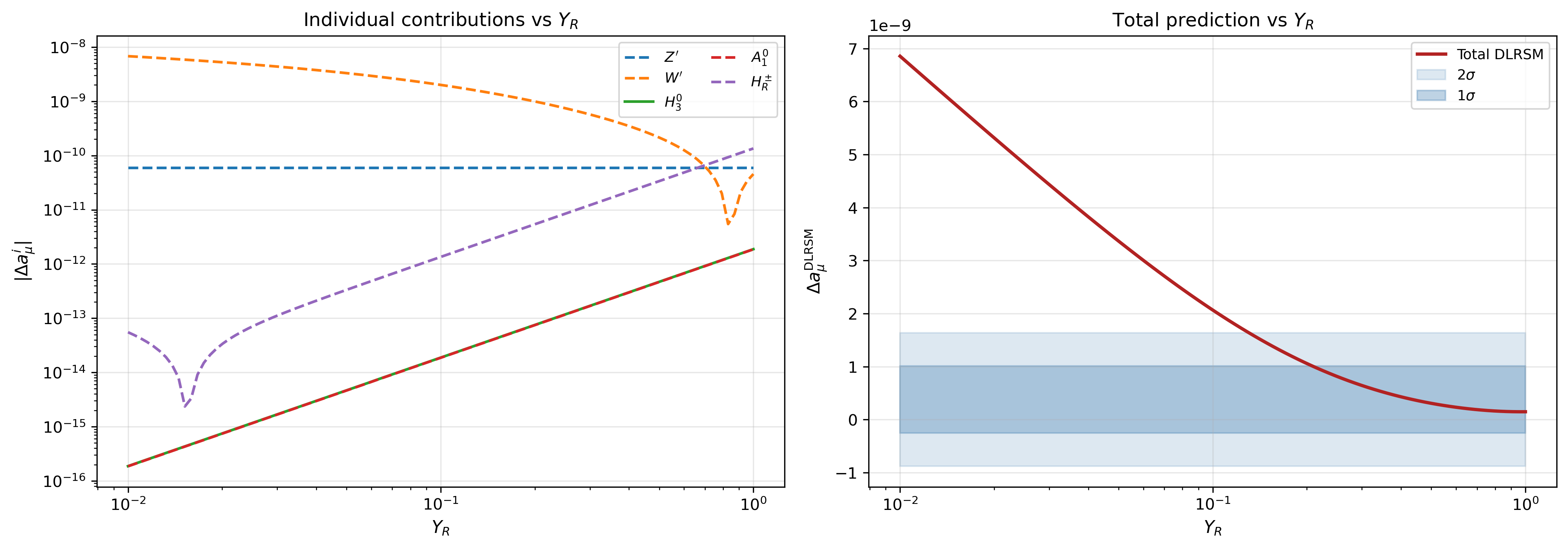}
\par\end{centering}
\centering{}\caption{Dependence of $\Delta a_{\mu}$on the Yukawa coupling $Y_{R}$, with
$v_{R}=10$ TeV, $\alpha_{23}=10^{-6}$, $\mu_{X}=1$ MeV. Left: Individual
contributions $|\Delta a_{\mu}^{i}|$. Right: Total prediction $\Delta a_{\mu}^{\text{DLRSM}}$
with experimental bands.}\label{fig:amu_YR}
\end{figure}

\begin{figure}
\begin{centering}
\includegraphics[scale=0.5]{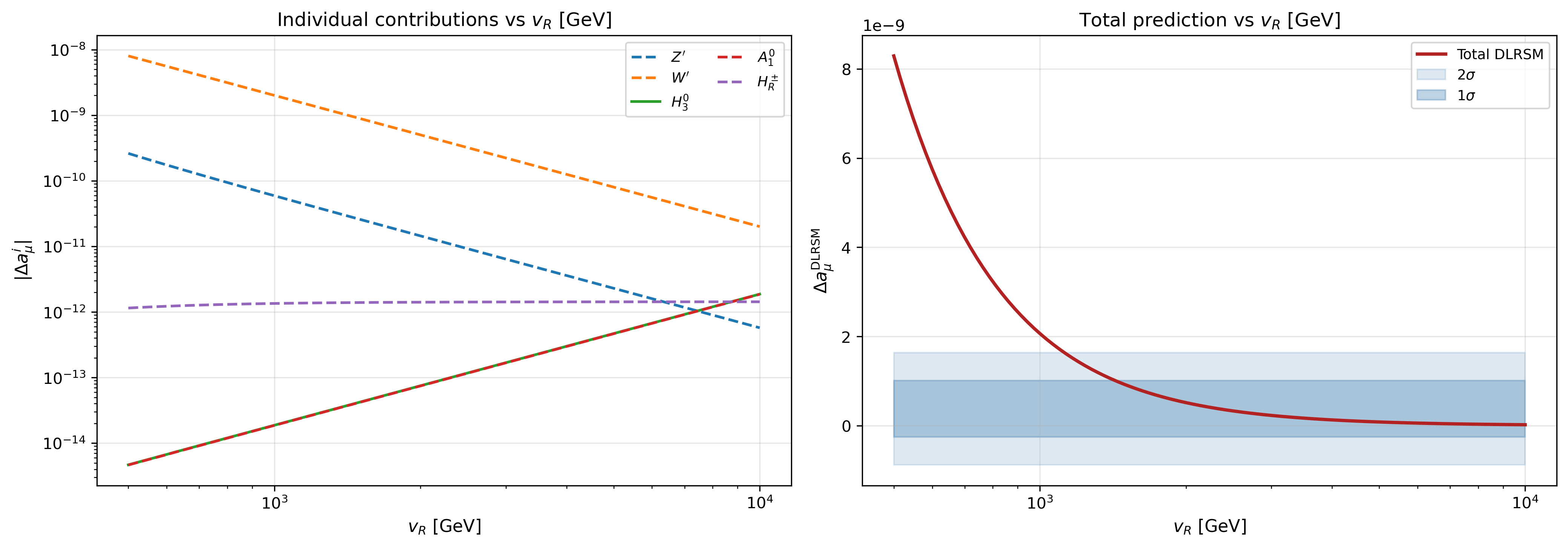}
\par\end{centering}
\centering{}\caption{Dependence of $\Delta a_{\mu}$ on the symmetry breaking scale $v_{R}$,
with $Y_{R}=0.1$, $\alpha_{23}=10^{-6}$, $\mu_{X}=1$ MeV. Left:
Individual contributions $|\Delta a_{\mu}^{i}|$. Right: Total prediction
$\Delta a_{\mu}^{\text{DLRSM}}$ compared to experimental bands.}\label{fig:amu_vR}
\end{figure}

In Figures \ref{fig:amu_YR} and \ref{fig:amu_vR} we vary $Y_{R}$
and $v_{R}$ parameter individually, respectively, while fixing the
remaining three to the reference values
\begin{align}
Y_{R}=0.1,\quad\alpha_{23}=10^{-6},\quad\mu_{X}=1\text{ MeV},\quad v_{R}=10\text{ TeV}.
\end{align}

On Figure \ref{fig:amu_YR}, the behavior of $a_{\mu}$ as a function
of $Y_{R}$ is shown. We observe that the most important contribution
to $a_{\mu}$ comes from $W^{\prime}$ in the left panel of Figure
\ref{fig:amu_YR}, the remaining contributions are negligible in comparison.
The total value of $\Delta a_{\mu}^{\text{DLRSM}}$ at low values
of $Y_{R}$ requires $Y_{R}\gtrsim10^{-1}$ to satisfy the experimental
$1\sigma$ band. 

Similarly, in left panel of Figure \ref{fig:amu_vR}, the individual
contributions to $a_{\mu}$are shown and we observe a similar behavior
to the case of $Y_{R}$. In this case, the $W^{\prime}$ contributions
is the most relevant contribution. The total $\Delta a_{\mu}^{\text{DLRSM}}$is
constrained to values of $v_{R}\gtrsim1$ TeV. 

\begin{figure}
\begin{centering}
\includegraphics[scale=0.28]{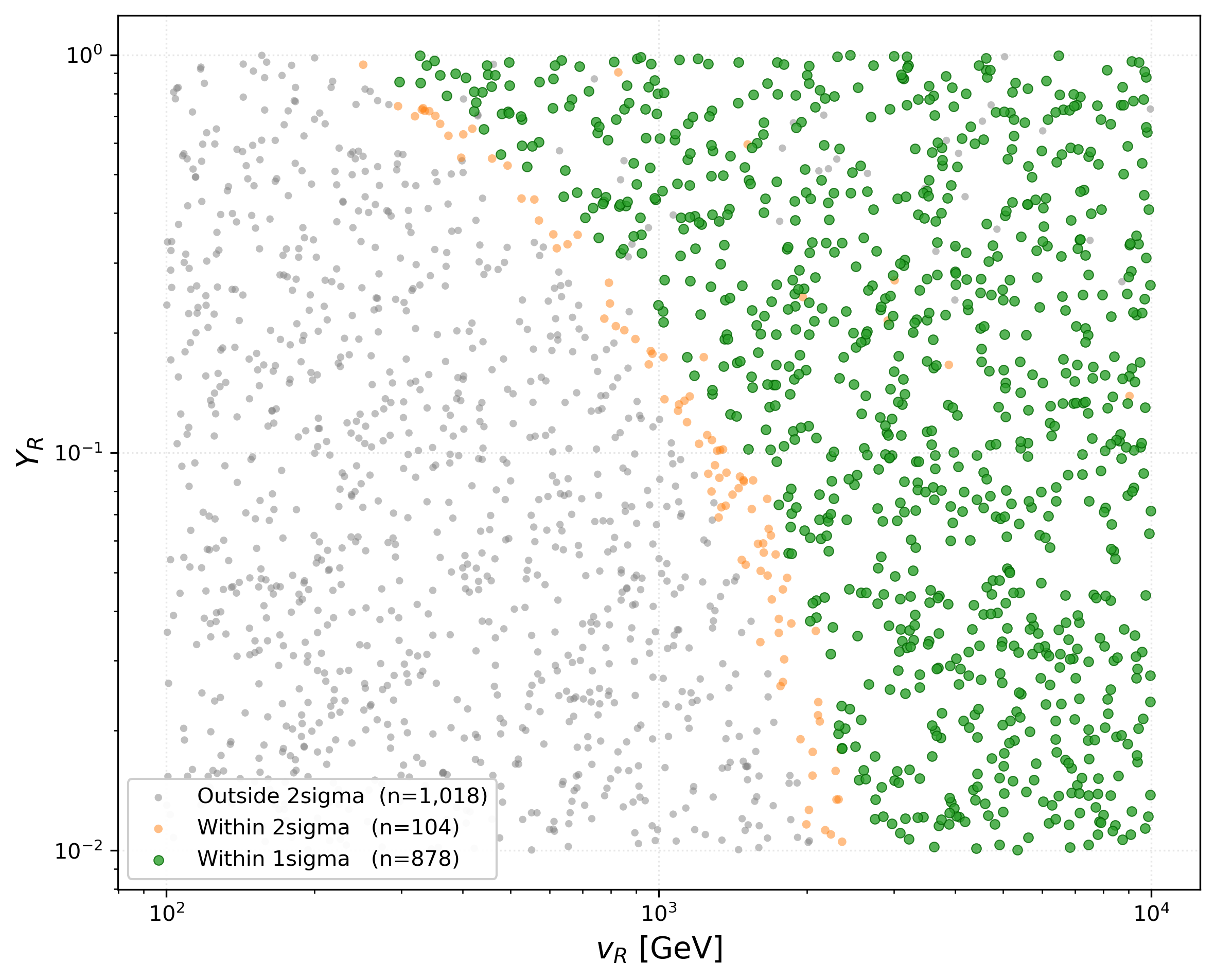}\includegraphics[scale=0.28]{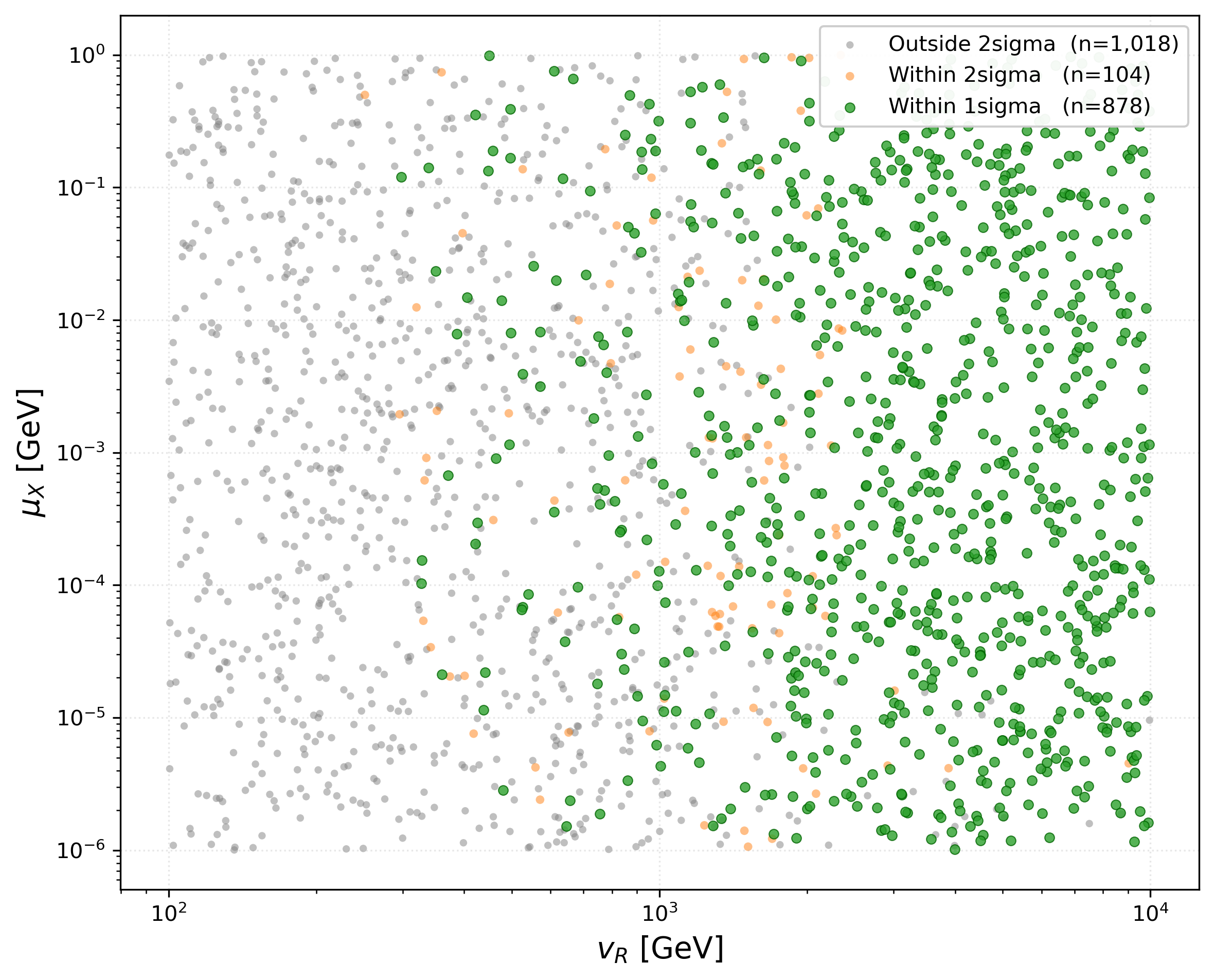}\includegraphics[scale=0.28]{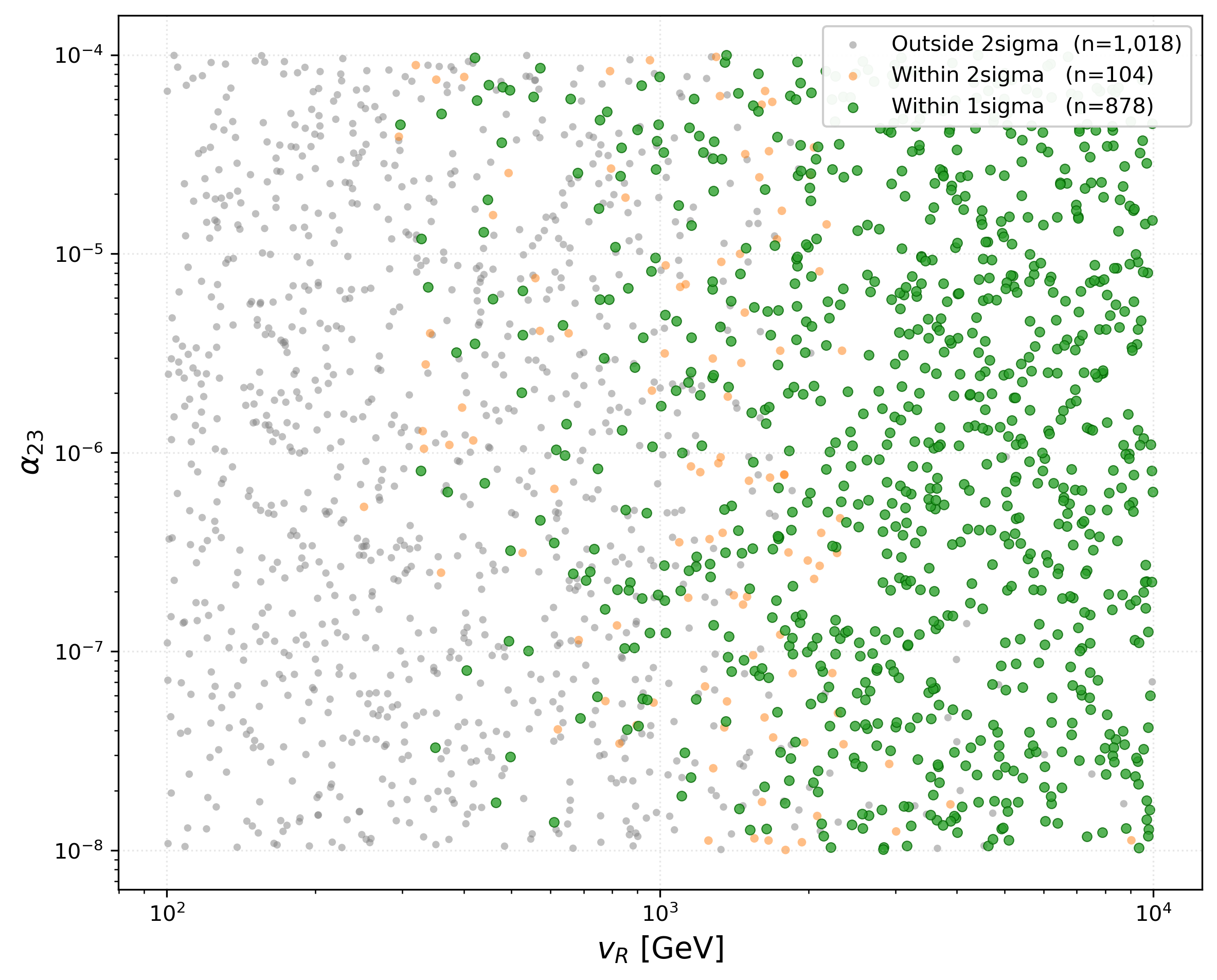}
\par\end{centering}
\caption{Results of a multi-parameter random scan projected onto the planes
of $(v_{R},Y_{R})$ (left), $(v_{R},\mu_{X})$ (center), and $(v_{R},\alpha_{23})$
(right). Points are color-coded based on their consistency with the
experimental measurement of $\Delta a_{\mu}$: green points lie within
the $1\sigma$ band, orange points within the $2\sigma$ band, and
gray points fall outside the $2\sigma$ allowed region. The clear
stratification demonstrates that the viability of the DLRSM solution
is primarily determined by the symmetry breaking scale $v_{R}$, favoring
a window starting at approximately $1$ TeV, largely independent of
the specific values of the other model parameters.}\label{fig:amu_region}
\end{figure}

To assess the robustness of the proposed solution across the full
multi-dimensional parameter space, we performed a random scan varying
the symmetry breaking scale $v_{R}$, the heavy neutrino Yukawa coupling
$Y_{R}$, the inverse seesaw parameter $\mu_{X}$, and the scalar
potential parameter $\alpha_{23}$ simultaneously, the range of the
scan are as follows
\begin{equation}
v_{R}\in[500,10^{4}]\text{ GeV},\quad Y_{R}\in[10^{-1},1],\quad\mu_{X}\in[10^{-6},1]\text{ GeV},\quad\alpha_{23}\in[10^{-8},10^{-4}].\label{eq:scan_ranges}
\end{equation}
The resulting viability regions are projected onto the $(v_{R},\mu_{X})$,
$(v_{R},Y_{R})$, and $(v_{R},\alpha_{23})$ planes in Figure \ref{fig:amu_region}.
A striking feature across all distributions is the sharp boundary
at $v_{R}\lesssim1$ TeV, below which the model generally overpredicts
the anomalous magnetic moment, resulting in contributions exceeding
the $2\sigma$ experimental upper bound (indicated by gray points).
Above this threshold, particularly for $v_{R}\gtrsim1$ TeV, we observe
a dense population of viable points (green) that satisfy the $1\sigma$
constraint. This viable window is remarkably insensitive to variations
in $\mu_{X}$ and $\alpha_{23}$ over several orders of magnitude,
confirming that the $W^{\prime}$ boson mass acts as the primary control
parameter. The projection onto the $(v_{R},Y_{R})$ plane plane further
shows that reducing $Y_{R}$ below $\sim0.1$ leads to an underprediction
of $\Delta a_{\mu}^{\text{DLRSM}}$, incompatible with the $1\sigma$
experimental band  

The analysis points toward new gauge boson masses $m_{W^{\prime}}\apprge325$
GeV, $m_{Z^{\prime}}\apprge385$ GeV in the MLRS condition and heavy
neutrinos $m_{n}\apprge700$ GeV. If we consider $g_{R}\neq g_{L}$,
the masses of new gauge bosons are as follows
\begin{equation}
M_{W^{\prime}}^{2}\approx\frac{1}{4}g_{R}^{2}v_{R}^{2},\quad M_{Z^{\prime}}^{2}\approx\frac{v_{R}^{2}}{4}\left(g_{R}^{2}+g_{BL}^{2}\right).
\end{equation}
In this case, the $a_{\mu}\left(W^{\prime}\right)$ it is not sensitive
to the values of $g_{R}$ as a consequence of the scaling of this
contribution
\begin{equation}
a_{\mu}\left(W^{\prime}\right)\propto\frac{g_{R}^{2}}{m_{W^{\prime}}^{2}}\approx\frac{1}{v_{R}^{2}},
\end{equation}
and we can conclude similarly that $v_{R}\gtrsim1$ TeV, if we consider
the perturbativy limit 
\begin{equation}
\frac{g_{R}^{2}}{4\pi}\apprle1,\label{eq:gR_perturvative}
\end{equation}

this points toward new gauge boson masses $m_{W^{\prime}}\apprge1625$
GeV, $m_{Z^{\prime}}\apprge1650$ GeV for $g_{R}=5g_{L}$, near to
the perturvative limit \eqref{eq:gR_perturvative}.

\section{Conclusions}\label{sec:Conclusions}

The experimental result for $\Delta a_{\mu}$, now consistent with
the Standard Model prediction within uncertainties, allows us to place
meaningful bounds on the parameter space of the DLRSM rather than
interpret the discrepancy as a signal of new physics. The dominant
new-physics contributions arise from the extended neutral and charged
gauge sectors, specifically from $Z'$ and $W^{\prime}$ loops, while
contributions from the new scalar sector are generally subdominant
except at small values of $\alpha_{23}$, where the charged scalar
$H_{R}^{\pm}$ become relevant. The parameter $\mu_{X}$, which controls
the small Majorana mass splitting in the inverse seesaw, is found
to have negligible impact on $\Delta a_{\mu}^{\text{DLRSM}}$, reflecting
the decoupling of the lightest neutrino mass scale from the observables
sensitive to the heavier spectrum. Our scan over the free parameters
establishes that $v_{R}\lesssim1$ TeV is excluded by the $g-2$ constraint
at the benchmark values considered, consequently, the analysis points
toward new gauge boson masses $m_{W^{\prime}}\apprge325$ GeV, $m_{Z^{\prime}}\apprge385$
GeV in the MLRS condition and heavy neutrinos $m_{n}\apprge700$ GeV.
In the case of $g_{R}\neq g_{L}$, the lower bounds could be stronger
such us $m_{W^{\prime}}\apprge1625$ GeV, $m_{Z^{\prime}}\apprge1650$
GeV. 
\begin{center}
\textbf{Acknowledgments} 
\par\end{center}

The research presented herein has been supported by the UNAM Postdoctoral
Program (POSDOC) and the PAPIIT project IN105825.

\appendix

\section{Neutral gauge boson matrix diagonalization}\label{sec:Gauge_diagonalization}

The neutral gauge boson mass matrix $M_{Z}^{2}$ in \eqref{eq:MZ_mass_matrix}
can be reduced to a block diagonal matrix by 
\begin{equation}
R=\begin{pmatrix}\sin\theta_{W} & -\cos\theta_{W} & 0\\
\sin\theta_{W} & \sin\theta_{W}\tan\theta_{W} & -\frac{\sqrt{\cos\left(2\theta_{W}\right)}}{\cos\theta_{W}}\\
\sqrt{\cos\left(2\theta_{W}\right)} & \sqrt{\cos\left(2\theta_{W}\right)}\tan\theta_{W} & \tan\theta_{W}
\end{pmatrix}\label{eq:R}
\end{equation}
with the following definitions
\begin{align}
e= & g\sin\theta_{W},\nonumber \\
\frac{1}{e^{2}}= & \frac{2}{g^{2}}+\frac{1}{g_{B-L}^{2}},\label{eq:ggBL_e_relations}
\end{align}
Then, the obtained block diagonal matrix
\begin{align*}
M_{0}^{2}= & R^{\top}M_{Z}^{2}R\\
= & \begin{pmatrix}0 & 0 & 0\\
0 & \frac{g^{2}\left(k_{1}^{2}+k_{2}^{2}\right)}{4\cos^{2}\theta_{W}} & -\frac{gg_{B-L}\left(k_{1}^{2}+k_{2}^{2}\right)}{2\cos\theta_{W}\tan\left(2\theta_{W}\right)}\\
0 & -\frac{gg_{B-L}\left(k_{1}^{2}+k_{2}^{2}\right)}{2\cos\theta_{W}\tan\left(2\theta_{W}\right)} & \frac{g^{2}v_{R}^{2}\cos^{2}\theta_{W}}{4\cos\left(2\theta_{W}\right)}+\frac{g^{2}\left(k_{1}^{2}+k_{2}^{2}\right)\cos\left(2\theta_{W}\right)}{4\cos^{2}\theta_{W}}
\end{pmatrix}
\end{align*}
which can be diagonalized by the rotation matrix $O\left(\zeta\right)$
over the angle $\zeta$ given by
\[
O\left(\zeta\right)=\begin{pmatrix}1 & 0 & 0\\
0 & \cos\left(\zeta\right) & \sin\left(\zeta\right)\\
0 & -\sin\left(\zeta\right) & \cos\left(\zeta\right)
\end{pmatrix},
\]
\begin{align}
\tan\left|2\zeta\right|\approx & \ensuremath{{\displaystyle \frac{4g_{B-L}\left(k_{1}^{2}+k_{2}^{2}\right)\cos\left(2\theta_{W}\right)}{gv_{R}^{2}\cos^{3}\theta_{W}\tan\left(2\theta_{W}\right)}}}.\label{eq:Z_mixing_angle}
\end{align}
In addition, with $R_{Z}=O\left(\zeta\right)R$ the neutral gauge
boson matrix \eqref{eq:MZ_mass_matrix}, is diagonalized by 
\begin{equation}
\hat{M}_{Z}^{2}=R_{Z}^{\top}M_{Z}^{2}R_{Z}.\label{eq:MZ_diagonal}
\end{equation}

\section{Master formulas for the one-loop contributions to $(g-2)_{\mu}$}\label{sec:amu_formulae}

This appendix collects the master integrals for one loop contributions
to $a_{\mu}=(g-2)_{\mu}/2$ following the notation of ~\cite{JEGERLEHNER20091}. 

\subsection{General structure and notation}

The generic one-loop correction to $a_{\mu}$from a diagram with internal
boson mass $M_{B}$, internal fermions mass $m_{F}$, couplings coefficients
$C_{X}$, takes the form
\begin{equation}
a_{\mu}=\frac{m_{\mu}^{2}}{8\pi^{2}M_{B}^{2}}\sum_{X}\left|C_{X}\right|^{2}\mathcal{L}_{X}\left(\epsilon,\lambda\right),\label{eq:amu_general}
\end{equation}

where the dimensionless parameters are
\[
\epsilon=\frac{m_{F}}{m_{\mu}},\quad\lambda=\frac{m_{\mu}}{M_{B}}.
\]

The index $X$ runs over the distinct coupling structures entering
the diagram $X=S,P,V,A$ and the loop function is given by
\begin{equation}
\mathcal{L}_{X}\left(\epsilon,\lambda\right)=\int_{0}^{1}\frac{Q_{X}\left(x,\epsilon,\lambda\right)}{\mathcal{D}\left(x,\epsilon,\lambda\right)}\,dx.\label{eq:loop_function}
\end{equation}
The denominator $\mathcal{D}$ from \eqref{eq:loop_function} depends
on whether the diagram has a neutral or charged boson exchange:
\begin{align}
\mathcal{D}_{\mathrm{neut}}\left(x,\epsilon,\lambda\right)= & \left(1-x\right)\left(1-\lambda^{2}x\right)+\epsilon^{2}\lambda^{2}x,\label{eq:Dneutral}\\
\mathcal{D}_{\mathrm{ch}}\left(x,\epsilon,\lambda\right)= & \left(\epsilon\lambda\right)^{2}\left(1-x\right)\left(1-\frac{x}{\epsilon^{2}}\right)+x.\label{eq:Dcharged}
\end{align}
 For the neutral exchange topologies (VFF and SFF), the numerator
$Q_{X}\left(x,\epsilon,\lambda\right)$ in \eqref{eq:loop_function}
is given by
\begin{align*}
Q_{S}^{\mathrm{neut}}= & x^{2}\left(1+\epsilon-x\right),\\
Q_{P}^{\mathrm{neut}}= & x^{2}\left(1-\epsilon-x\right),\\
Q_{V}^{\mathrm{neut}}= & 2x\left(1-x\right)\left(x-2\left(1-\epsilon\right)\right)+\lambda^{2}\left(1-\epsilon\right)^{2}x^{2}\left(1+\epsilon-x\right),\\
Q_{A}^{\mathrm{neut}}= & 2x\left(1-x\right)\left(x-2\left(1+\epsilon\right)\right)+\lambda^{2}\left(1+\epsilon\right)^{2}x^{2}\left(1-\epsilon-x\right).
\end{align*}
For the charged exchange topologies (FVV and FSS):
\begin{align*}
Q_{S}^{\mathrm{ch}}= & -x\left(1-x\right)\left(x+\epsilon\right),\\
Q_{P}^{\mathrm{ch}}= & -x\left(1-x\right)\left(x-\epsilon\right),\\
Q_{V}^{\mathrm{ch}}= & 2x^{2}\left(1+x-2\epsilon\right)+\lambda^{2}\left(1-\epsilon\right)^{2}x\left(1-x\right)\left(x+\epsilon\right),\\
Q_{A}^{\mathrm{ch}}= & 2x^{2}\left(1+x+2\epsilon\right)+\lambda^{2}\left(1+\epsilon\right)^{2}x\left(1-x\right)\left(x-\epsilon\right).
\end{align*}
So, VFF (V neutral boson) contribution to $a_{\mu}$ is given by
\[
a_{\mu}\left(\text{VFF}\right)=\frac{m_{\mu}^{2}}{8\pi^{2}M_{B}^{2}}\left[\left|C_{V}\right|^{2}\mathcal{L}_{V}^{neut}\left(\epsilon,\lambda\right)+\left|C_{A}\right|^{2}\mathcal{L}_{A}^{neut}\left(\epsilon,\lambda\right)\right]
\]
with 
\[
\mathcal{L}_{V,A}^{neut}\left(\epsilon,\lambda\right)=\int_{0}^{1}\frac{Q_{V,A}^{neu}\left(x,\epsilon,\lambda\right)}{\mathcal{D}^{neu}\left(x,\epsilon,\lambda\right)}\,dx.
\]
The topology assignment formulae is summarized in Table \ref{tab:Topologies_formulae}.
\begin{table}
\begin{centering}
\begin{tabular}{|c|c|c|c|}
\hline 
$\Theta$ & Figure & Numerator & Denominator\tabularnewline
\hline 
\hline 
VFF & \ref{fig:Feynman_FXX}(Left) & $Q_{V}^{\mathrm{neu}},\,Q_{A}^{\mathrm{neu}}$ & $\mathcal{D}_{\mathrm{neut}}$\tabularnewline
\hline 
SFF & \ref{fig:Feynman_XFF}(Left) & $Q_{S}^{\mathrm{neu}},\,Q_{P}^{\mathrm{neu}}$ & $\mathcal{D}_{\mathrm{neut}}$\tabularnewline
\hline 
FVV & \ref{fig:Feynman_FXX}(Right) & $Q_{V}^{\mathrm{ch}},\,Q_{A}^{\mathrm{ch}}$ & $\mathcal{D}_{\mathrm{ch}}$\tabularnewline
\hline 
FSS & \ref{fig:Feynman_XFF}(Right) & $Q_{S}^{\mathrm{ch}},\,Q_{P}^{\mathrm{ch}}$ & $\mathcal{D}_{\mathrm{ch}}$\tabularnewline
\hline 
\end{tabular}
\par\end{centering}
\caption{Topology assignment to each contribution to $a_{\mu}$ in the unitary
gauge~\cite{JEGERLEHNER20091}}\label{tab:Topologies_formulae}

\end{table}

\bibliographystyle{unsrt}
\bibliography{biblioteca}

\end{document}